\documentclass[preprint, showpacs,preprintnumbers,amsmath,amssymb,nofootinbib]{revtex4}
\usepackage{mathrsfs}
\usepackage{amsmath}
\usepackage{amsfonts}
\usepackage{latexsym}
\usepackage{amsfonts}
\usepackage{graphicx}
\usepackage{epsf}
\usepackage{dcolumn}
\usepackage{bm}

\newcommand{\bea}[1]{\begin{eqnarray}\label{#1}}
 \newcommand{\eea}{\end{eqnarray}}

 \def\gsim{ \lower .75ex \hbox{$\sim$} \llap{\raise .27ex \hbox{$>$}} }
 \def\lsim{ \lower .75ex \hbox{$\sim$} \llap{\raise .27ex \hbox{$<$}} }
\def\/{\over}

\begin{document}

\title{\bf The stability of Einstein static universe in the DGP braneworld}

\author{ Kaituo Zhang, Puxun Wu, Hongwei Yu \footnote{Corresponding
author}}
\address{Department of Physics and Institute of  Physics,\\ Hunan Normal
University, Changsha, Hunan 410081, China \\
Key Laboratory of Low Dimensional Quantum Structures and Quantum
Control of Ministry of Education, Hunan Normal University, Changsha,
Hunan 410081, China}

\begin{abstract}
The stability of an Einstein static universe in the DGP braneworld
scenario is studied in this letter. Two separate branches denoted by
$\epsilon=\pm1$ of  the DGP model are analyzed. Assuming the
existence of a perfect fluid with a constant equation of state, $w$,
in the universe, we find that, for the branch with $\epsilon=1$,
there is no a stable Einstein static solution, while, for the case
with $\epsilon=-1$, the Einstein static universe exists and it is
stable when $-1<w<-\frac{1}{3}$.  Thus, the universe can stay at
this stable state past-eternally and may undergo a series of
infinite, non-singular oscillations.   Therefore, the big bang
singularity problem in the standard cosmological model can be
resolved.

\end{abstract}


\maketitle
\section{Introduction}

 Although most of the problems in the standard cosmological model can be resolved by the inflation
theory, the resolution of the existence of a big bang singularity in
the early universe is still elusive. Based upon the string/M-theory,
the pre-big bang~\cite{Gasperini2003}  and
ekpyrotic/cyclic~\cite{Khoury2001} scenarios have been proposed to
address the issue. Recently, Ellis et al proposed, in the context of
general relativity, a new scenario, called an emergent
universe~\cite{Ellis2004a, Ellis2004b} to avoid this singularity. In
this scenario, the space curvature is positive, which is supported
by the recent observation from WMAP7~\cite{Komatsu2010} where it was
found that a closed universe is favored  at the $68\%$ confidence
level,  and the universe stays, past-eternally, in an Einstein
static state and then evolves to a subsequent inflationary phase.
So, in an emergent theory, the universe originates from an Einstein
static state rather than from a big bang singularity.  However, the
Einstein static universe in the classical general relativity is
unstable, which means that it is extremely difficult for  the
universe  to remain in such an initial static state in a long time
due to the existence of perturbations, such as the quantum
fluctuations. Therefore,  the original emergent model does not seem
to resolve the big bang singularity problem successfully as
expected.

However, in the early epoch, the universe is presumably under
extreme physical conditions, the  realization of the initial state
may be affected by  novel physical effects, such as those stemming
from quantization of gravity, or a modification of general
relativity or even other new physics. As a result, the stability of
the Einstein static state has been examined in various
cases~\cite{Carneiro2009, Mulryne2005, Parisi2007, Wu2009,
Lidsey2006, Bohmer2007, Seahra2009, Bohmer2009, Barrow2003,
Barrow2009, Clifton2005,Boehmer2010, Boehmer20093, Wu20092,
Odrzywolek2009},  from loop quantum gravity ~\cite{Mulryne2005,
Parisi2007, Wu2009} to  modified gravity (for a review see
Ref.~\cite{Boehmer2010} ), from Horava-Lifshitz
gravity~\cite{Boehmer20093, Wu20092} to Shtanov-Sahni braneworld
scenario~\cite{Lidsey2006}

In this paper,  we plan to examine the stability of the Einstein
static universe in the DGP brane-world model~\cite{Dvali2000}.
In this braneworld, the whole energy-momentum is confined on a three
dimensional brane embedded in a five-dimensional, infinite-volume
Minkowski bulk. Since there are two different ways to embed the
4-dimensional brane  into the 5-dimensional spacetime, the DGP model
has two separate branches denoted by $\epsilon$ with distinct
features. The $\epsilon=+1$ branch  can explain the present
accelerating cosmic expansion without the introduction of dark
energy~\cite{Deffayer2001}, while for the $\epsilon=-1$ branch, dark
energy is needed in order to yield an accelerating expansion, as is
the case in the LDGP model~\cite{Lue2004} and the QDGP
model~\cite{Chimento2006}. Using the $H(z)$, CMB shift and Sne Ia
observational data,  Lazkoz and Majerotto~\cite{Lazkoz2007} found
that the LDGP and QDGP are slightly more favored than the
self-accelerating DGP model. Let us also note that a crossing of a
phantom divide line, which is favored by the recent various
observational data~\cite{Alam2004}, is possible with a single scalar
field~\cite{Zhang20062, Chimento2006}  in the $\epsilon=-1$ branch.
In addition, inflation in the DGP  model displays some new
characteristics. It should be noted, however, that only in the
$\epsilon=-1$ case can inflation exit
spontaneously~\cite{Bouhmadi-Lopez2004, Cai2004,
Papantonopoulos2004, Zhang2004, Zhang2006, Campo2007}. Also, in
contrast to the Randall-Sundrum~\cite{Randall1999} and
Shtanov-Sahni~\cite{Shtanov2003} braneworld scenarios with high
energy modifications to general relativity, the DGP brane produces a
low energy modification (for a review of the phenomenology of the
DGP model, see Ref.~\cite{Lue2006}).

\section{The friedmann equation in DGP braneworld}

For a homogeneous and isotropic universe which is described by the
Friedmann-Robertson-Walker (FRW) metric.
\begin{eqnarray}
ds^2=-dt^2+a^2(t)\bigg(\frac{dr^2}{1-kr^2}+r^2d^2\Omega\bigg)\;,
\end{eqnarray}
the Friedmann equation on the warped DGP brane can  be written
as~\cite{Maeda2003}
\begin{eqnarray}
H^2+\frac{k}{a^2}=\frac{1}{3\mu^2}[\rho+\rho_0(1+\epsilon
\mathcal{A}(\rho,a))]\;,
\end{eqnarray}
where $H$ is the Hubble parameter, $k$ is the constant curvature of
the three-space of the FRW metric, $\rho$ is the total energy
density and  $\mu$ is a  parameter denoting the strength of the
induced gravity  on the brane. 
For $\epsilon=-1$, the brane tension can be assumed to be positive,
while for $\epsilon=+1$, it is negative. $\mathcal{A}$ is given by
\begin{eqnarray}
\mathcal{A}=\bigg[\mathcal{A}_0^2+\frac{2\eta}{\rho_0}\bigg(\rho-\mu^2\frac{\mathcal
{E}_0}{a^4}\bigg)\bigg]^{1/2}\;,
\end{eqnarray}
where \begin{eqnarray}\mathcal{A}_0=\sqrt{1-2\eta
\frac{\mu^2\Lambda}{\rho_0}},\quad \eta=\frac{6m_5^6}{\rho_0\mu^2}
\;\;\; (0<\eta\leq 1), \quad
\rho_0=m_\lambda^4+6\frac{m_5^6}{\mu^2}\;,\end{eqnarray} with
$\Lambda$ defined as
\begin{eqnarray}\Lambda=\frac{1}{2}(^{(5)}\Lambda+\frac{1}{6}\kappa_5^6\lambda^2)\;.
\end{eqnarray}
Here $\kappa_5$ is the 5-dimensional Newton constant,
$^{(5)}\Lambda$ the 5-dimensional cosmological constant in the bulk,
$\lambda$  the brane tension, and $\mathcal {E}_0$  a constant
related to Weyl radiation. For simplicity, we will neglect the dark
radiation term and restrict ourselves to the Randall- Sundrum
critical case, i.e. $\Lambda=0$, then Eq.(2)  simplifies to
\begin{eqnarray}
H^2+\frac{k}{a^2}=\frac{1}{3\mu^2}\bigg(\rho+\rho_0+\epsilon\rho_0\sqrt{1+\frac{2\eta\rho}{\rho_0}}\bigg).
\end{eqnarray}
Since in the very early era of the universe, the total energy
density should be very high. Thus, we will, in the following, only
consider the ultra high energy limit, $\rho$$\gg$$\rho_0$. In
addition, we are interested in a closed universe, so we set the
constant curvature $k$ to be $+1$. As a result,  the Friedmann
equation reduces to
\begin{eqnarray}\label{FEq}
H^2=\frac{1}{3\mu^2}(\rho+\epsilon\sqrt{2\rho\rho_{0}})-\frac{1}{a^2}.
\end{eqnarray}
This describes a 4-dimensional gravity with minor corrections, which
implies that $\mu$ must have an energy scale as the Planck scale in
the DGP model.

The energy density $\rho$ of a perfect fluid in the universe
satisfies the conservation equation
\begin{eqnarray}
\dot{\rho}=-3H(1+w)\rho,
\end{eqnarray}
where $w=\frac{p}{\rho}$ is the equation of state of the perfect
fluid.  A constant $w$ is considered in the present paper,  which is
a good approximation if the perfect fluid is a scalar field and the
variation of the potential of scalar field is very slow with time. 

Differentiating Eq.~(\ref{FEq}) with respect to cosmic time, one
gets
\begin{eqnarray}
\dot{H}=-\frac{1}{2\mu^2}(\rho+p)\bigg(1+\epsilon\sqrt{\frac{\rho_0}{2\rho}}\bigg)+\frac{1}{a^2},
\end{eqnarray}
Combining this equation with the Friedmann equation given in
Eq.~(\ref{FEq}), we have
\begin{eqnarray}\label{dEq}
\frac{\ddot{a}}{a}=-\frac{1}{2\mu^2}(\rho+p)\bigg(1+\epsilon\sqrt{\frac{\rho_0}{2\rho}}\bigg)+\frac{1}{3u^2}(\rho+\epsilon\sqrt{2\rho\rho_0}).
\end{eqnarray}

\section{The Einstein static solution}
The Einstein static solution is given by the conditions $\dot{a}=0$
and $\ddot{a}=0$,  which imply
\begin{eqnarray}a=a_{Es},\qquad
H(a_{Es})=0\;.
\end{eqnarray}
At the critical point determined by above conditions,  we find,
using  Eq.~(\ref{dEq})
\begin{eqnarray}\label{rhoEs}
\sqrt{\rho_{Es}}=\frac{\epsilon\sqrt{2\rho_0}(1-3\omega)}{2(1+3\omega)},
\end{eqnarray}
which means that in this dynamical system, there is only one
Einstein static state solution. In order to guarantee the physical
meaning of $\rho_{Es}$, it is necessary that
\begin{eqnarray}\frac{\epsilon(1-3\omega)}{1+3\omega}\geq
0.\end{eqnarray}

Substituting Eq.~(\ref{rhoEs}) into the Friedmann equation, we
obtain at the critical point
\begin{eqnarray}\label{aEs}
\frac{1}{a^2_{Es}}=\frac{\rho_0(1-3\omega)(1+\omega)}{2\mu^2(1+3\omega)^2},
\end{eqnarray}
with the requirement $(1-3\omega)(1+\omega)>0$.

Before analyzing the stability of the critical point, we want to
express Eq.~(\ref{dEq}) in terms of $a$ and $H$. To do so, we  put
the Friedmann equation in a different way
\begin{eqnarray}
 \sqrt{\rho}=\frac{\sqrt{2}}{2}\bigg(-\epsilon\sqrt{\rho_0}+\sqrt{\rho_0+6\mu^2\bigg(H^2+\frac{1}{a^2}\bigg)}\bigg).
\end{eqnarray}
Thus Eq.~(\ref{dEq}) can be re-written as
\begin{eqnarray}
\frac{\ddot{a}}{a}&=&-\frac{1}{4\mu^2}(1+\omega)\rho_o+\frac{1}{4\mu^2}\epsilon(1+\omega)\sqrt{\rho_0^2+6\mu^2\rho_0\bigg(H^2+\frac{1}{a^2}\bigg)}\nonumber\\
&&\quad-\frac{1}{2}(1+3\omega)\bigg(H^2+\frac{1}{a^2}\bigg).
\end{eqnarray}

Now we study the stability of the critical point. For convenience,
we introduce two  variables
\begin{eqnarray}
x_1=a,\quad x_2=\dot{a}.
\end{eqnarray}
It is then easy to obtain the following equations
\begin{eqnarray}
\dot{x_1}=x_2,
\end{eqnarray}
\begin{eqnarray}
\dot{x_2}=-\frac{1}{4\mu^2}\rho_0(1+w)x_1+\frac{1}{4\mu^2}\epsilon(1+\omega)\sqrt{\rho_0^2x_1^2+6\mu^2\rho_0(1+x_2^2)}-\frac{1}{2}(1+3\omega)\frac{x_2^2+1}{x_1}.
\end{eqnarray}
In these variable, the Einstein static solution corresponds  to the
fixed point, $x_1=a_{Es},\; x_2=0$. The stability of the critical
point is determined by the eigenvalue of the coefficient matrix
resulting from linearizing the system described by above two
equations near the critical point. Using $\lambda^2$ to denote the
eigenvalue, we have
\begin{eqnarray}
\lambda^2=\frac{\rho_0}{8\mu^2}\epsilon(1+\omega)|1+3\omega|-\frac{3\rho_0}{2\mu^2}\frac{\omega(1+\omega)}{1+3\omega}
\end{eqnarray}
If $\lambda^2<0$, the corresponding equilibrium point is a center
point otherwise it is a saddle one. In order to analyze the
stability of the critical point in detail,  we now divide our
discussions into two cases, i.e., $\epsilon=-1$ and $\epsilon=1$.

\vspace{0.5cm} \textbf{A}. $\epsilon=1$ \vspace*{0.5cm}

In this case, $-\frac{1}{3}<\omega<\frac{1}{3}$  is required  to
ensure that the critical point is physically meaningful. It then
follows that $\lambda^2>0$, which means that this critical point is
a saddle point. Thus, there is no stable Einstein static solution,
and an emergent universe is not realistic in this case.

\vspace{0.5cm} \textbf{B}. $\epsilon=-1$  \vspace{0.5cm}

\begin{figure}[htbp]
\includegraphics[width=7cm]{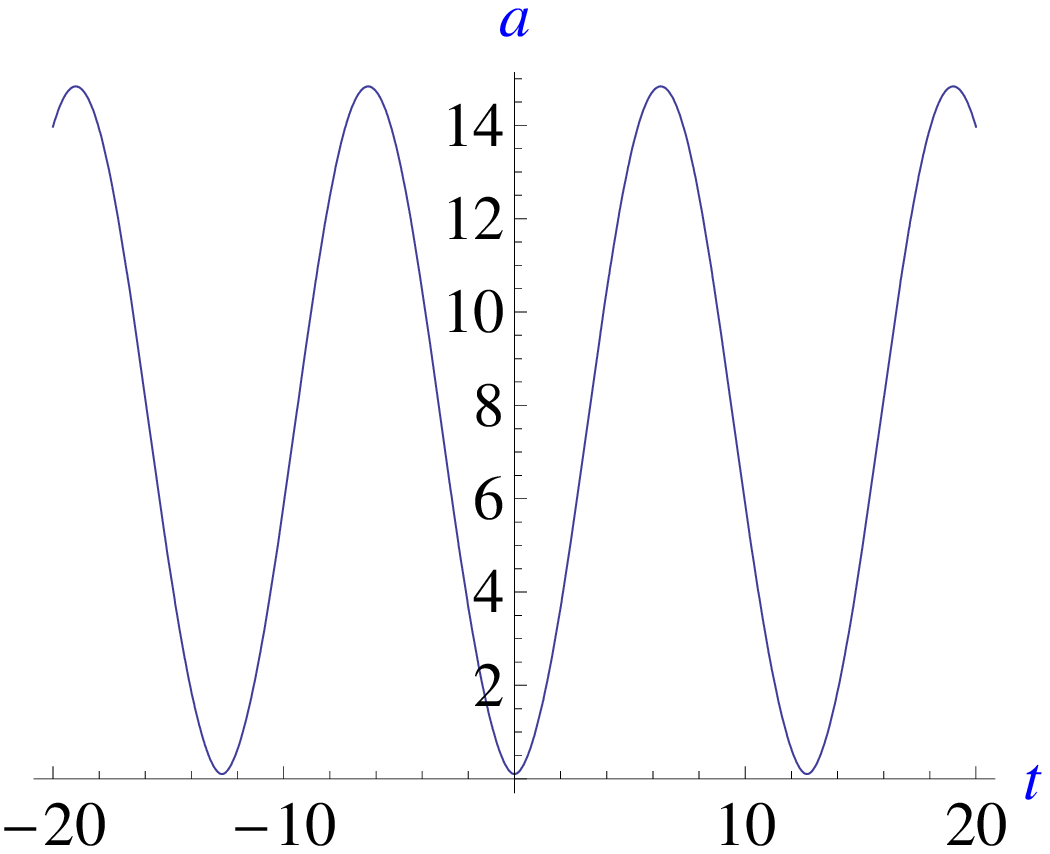}\quad\includegraphics[width=7cm]{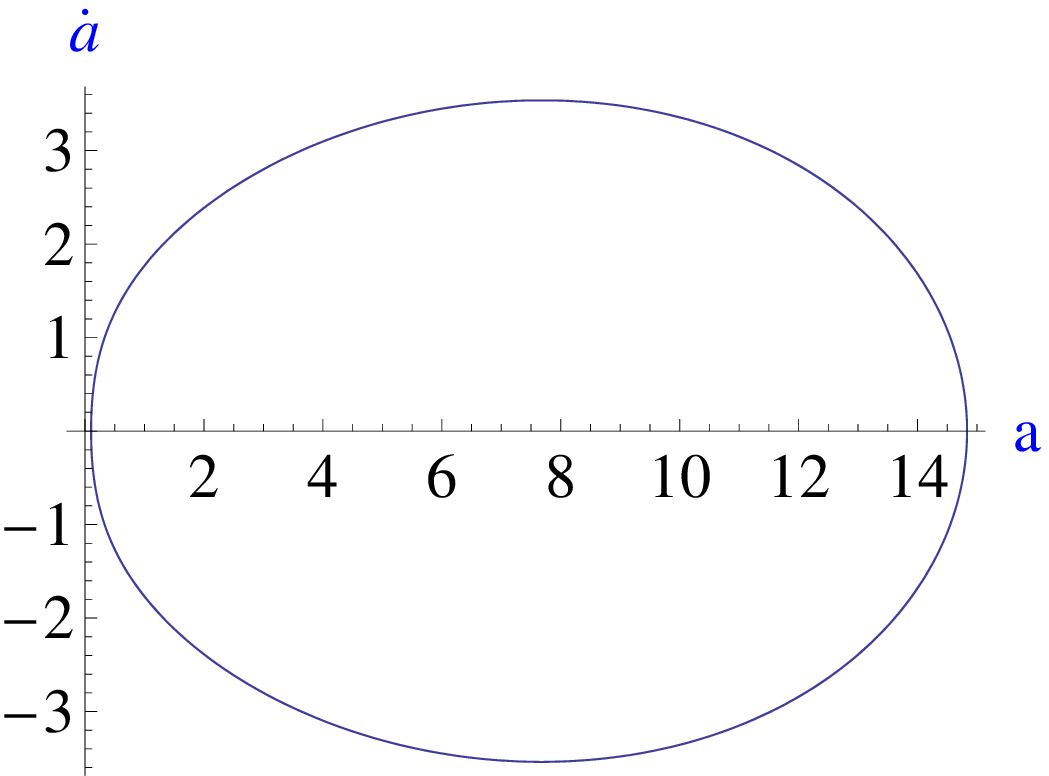}
 \caption{\label{Fig1} The evolutionary curve of the scale factor with time (left) and
the phase diagram  in space ($a$, $\dot{a}$) (right) for the case
$\epsilon=-1$ in Planck unit  and with $w = -0.50$. }
\end{figure}

Now the requirement for the critical point to be physically
meaningful is $-1<\omega<-\frac{1}{3}$. This exactly agrees with the
condition of stability ($\lambda^2<0$).  Hence,  as long as the
critical point exists, it is always stable. So, if the scale factor
satisfies the condition given in Eq.~(\ref{aEs}) initially and $w$
is within the region of stability, the universe can stay at this
stable state past-eternally, and may undergo a series of infinite,
non-singular oscillations, as shown in Fig.~(\ref{Fig1}). As a
result, the big bang singularity can be avoided successfully.

\section{Leaving the Einstein static state}
Now, we have shown that an stable Einstein static state exists in
the $\epsilon=-1$ branch. However,  in order to have a successful
cosmological scenario, a graceful exit to an inflationary epoch is
needed. This is possible in the following sense.  In the analysis
carried out in the present paper,  the equation of state $w$ of the
perfect fluid in the universe is assumed to be a constant, and this
is a good approximation if the energy component in the early
universe is only that of a minimally coupled scalar field with a
self-interaction potential. One can show that the kinetic energy and
potential energy of this scalar field should be both non-zero
constants for an Einstein static solution~\cite{Ellis2004a,
Ellis2004b, Mulryne2005, Lidsey2006}. That is to say, the scalar
field rolls along a plateau potential. However a realistic
inflationary model clearly requires the potential to vary as the
scalar field evolves. Thus,  the constant potential is   merely a
past-asymptotic limit of a smoothly varying one, as pointed out in
Refs.~\cite{Ellis2004a, Ellis2004b, Mulryne2005, Lidsey2006,
Carneiro2009}. So, the essentially slowly varying potential will
eventually break the equilibrium of the Einstein static state and
lead to  an exit from the initial Einstein phase to an inflationary
one. Some specific forms of such a potential that implements these
features have been constructed in Refs~\cite{Ellis2004a, Ellis2004b,
Mulryne2005}.

\section{Conclusions}

In this paper, we have studied the existence and stability of the
Einstein static universe in the DGP braneworld scenario. By assuming
the existence of a perfect fluid with a constant equation of state,
which is a good approximation if the perfect fluid is a scalar field
and the variation of the potential of scalar field is very slow with
time, we have shown that for the branch with $\epsilon=1$, there is
no stable Einstein static universe, whereas, for the branch with
$\epsilon=-1$, the Einstein static universe exists and it is stable
if the equation of state $w$ satisfies $-1<\omega<-\frac{1}{3}$.
Thus, the universe can stay at this stable state past-eternally, and
may undergo a series of infinite, non-singular oscillations. Hence,
in the $\epsilon=-1$ branch of the DGP model, the universe can
originate from an Einstein static state and then enter an inflation
era. Furthermore, the universe can exit, spontaneously,  this
inflation phase
 to a radiation dominated era, as shown in previous studies~\cite{Bouhmadi-Lopez2004, Cai2004,
Papantonopoulos2004, Zhang2004, Zhang2006, Campo2007}. As a result,
the big bang singularity problem in the standard cosmological
scenario can be resolved successfully.

\begin{acknowledgments}

This work was supported in part by the National Natural Science
Foundation of China under Grants Nos. 10775050, 10705055 and
10935013, the SRFDP under Grant No. 20070542002,  the FANEDD under
Grant No. 200922, the National Basic Research Program of China under
Grant No. 2010CB832803, the NCET under Grant No.09-0144, the PCSIRT under Grant No. IRT0964, and the Programme for the Key Discipline in
Hunan Province.

\end{acknowledgments}


\end{document}